\documentclass[preprint,showpacs,preprintnumbers,prb,superscriptaddress]{revtex4}
\usepackage{graphicx}
\usepackage{dcolumn}
\usepackage{bm}

\providecommand{\func}[1]{\mathrm{#1}}

\begin{document}

\title{First-Principles Study of Correlation Effects in VO$_{2}$: Peierls
vs. Mott-Hubbard}
\author{R. Sakuma}
\email{reis@faculty.chiba-u.jp}
\affiliation{Research Institute for Computational Sciences, National Institute of
Advanced Industrial Science and Technology, Tsukuba, Ibaraki 305-8568, Japan}
\affiliation{Japan Science and Technology Agency, CREST, Kawaguchi, Saitama 332-0012,
Japan}
\affiliation{Graduate School of Advanced Integration Science, Chiba University, Chiba 263-8522, Japan}
\author{T. Miyake}
\affiliation{Research Institute for Computational Sciences, National Institute of
Advanced Industrial Science and Technology, Tsukuba, Ibaraki 305-8568, Japan}
\affiliation{Japan Science and Technology Agency, CREST, Kawaguchi, Saitama 332-0012,
Japan}
\author{F. Aryasetiawan}
\affiliation{Research Institute for Computational Sciences, National Institute of
Advanced Industrial Science and Technology, Tsukuba, Ibaraki 305-8568, Japan}
\affiliation{Japan Science and Technology Agency, CREST, Kawaguchi, Saitama 332-0012,
Japan}
\affiliation{Graduate School of Advanced Integration Science, Chiba University, Chiba 263-8522, Japan}

\date{\today}

\begin{abstract}
We present a study of VO$_{2}$ in the rutile and monoclinic ($M_{1}$) phases
by means of all-electron full-potential LMTO GW calculation. Full frequency
dependence and the off-diagonal matrix elements of the self-energy are taken
into account. As a result of dynamical correlation, a satellite structure is
found above the $t_{2g}$ quasiparticle peak but not below, in both the
rutile and monoclinic phases. For the monoclinic structure, the insulating
state is not obtained within the usual 1-shot GW calculation. We perform a
simplified ``self-consistent'' GW scheme by adding a uniform shift to the
conduction band levels and recalculating the quasiparticle wavefunctions
accordingly. An insulating solution with a gap of approximately 0.6 eV is
obtained, in agreement with experiments.
\end{abstract}

\maketitle

\section{Introduction}

Vanadium dioxide undergoes a metal-insulator transition (MIT) at the
transition temperature $\approx 340$ K\cite{Morin}, and much work has been
done both experimentally and theoretically to investigate the role of
electron correlations in this transition. At high temperature VO$_{2}$ is
metallic and forms tetragonal rutile (R) structure with space group $%
P4_{2}/mnm$, while at low temperature VO$_{2}$ forms monoclinic ($M_{1}$)
structure with space group $P2_{1}/c$, and becomes insulating with a band
gap of $0.6$ eV \cite{Koethe}. The striking feature of $M_{1}$ phase is the
dimerization of vanadium atoms with the zigzag-like displacements, which
leads to a doubling of the unit cell along the $c$ axis. In both phases the
crystal field splits vanadium $t_{2g}$ bands into a doubly degenerate $%
e_{g}^{\pi }$ band and a non-degenerate $a_{1g}$ band. Goodenough \cite%
{Goodenough} proposed a Peierls (band-like) picture of the insulating state:
In the metallic VO$_{2}$, the $a_{1g}$ band overlaps with the $e_{g}^{\pi }$
band, and both bands are partially filled. In the insulating VO$_{2}$, the
pairing of vanadium atoms leads to the bonding-antibonding splitting of two $%
a_{1g}$ bands, which separates the (bonding) $a_{1g}$ band and $e_{g}^{\pi }$
band. The bonding $a_{1g}$ band becomes filled, while the $e_{g}^{\pi }$
band becomes empty, and the gap opens up between them. This scenario is also
supported by Wentzcovitch \textit{et al} \cite{Wentzcovitch}, who showed
that first-principles molecular dynamics calculations within the local
density approximation (LDA) of density functional theory can reproduce the
structure and energy difference of these two phases in good accuracy. They
failed to reproduce an insulating monoclinic VO$_{2}$, but they attributed
this failure to the problem of the LDA. %
However, since this Peierls model cannot explain the metastable insulating
state of another monoclinic ($M_{2}$) structure \cite{Pouget1,Pouget2},
where only half of vanadium atoms are dimerized, the importance of
electron-electron correlations was emphasized \cite{Zylbersztejn,Rice}. To
study the effect of correlations beyond the LDA, theoretical works using LDA
+ DMFT (dynamical mean field theory \cite{Georges}) have recently been done%
\cite{Liebsch,Biermann,Laad}, but a consensus has not been reached. Liebsch 
\textit{et al}. performed a single-site multiband DMFT calculation with the
quantum Monte Carlo method, but their LDA+DMFT result does not reproduce an
insulating VO$_{2}$ \cite{Liebsch}. On the other hand, Laad \textit{et al}. 
concluded from the LDA+DMFT calculation within the iterative perturbation
theory (IPT), that the MIT is of the Mott-Hubbard type \cite{Laad}. To
account for the spacial correlation between dimerized vanadium atoms,
Biermann \textit{et al}. performed cluster-DMFT calculations in this system
and succeeded in reproducing the two phases \cite{Biermann}. In their result
for insulating VO$_{2}$ the antibonding $a_{1g}$ band is strongly
renormalized to form the upper Hubbard band due to correlation, but the peak
of the spectral function just below the Fermi energy is, interestingly, not
an incoherent lower Hubbard band but a quasiparticle peak of the $a_{1g}$
band, which indicates insulating VO$_{2}$ has also a band-like nature.

The above DMFT model approaches are parameter-dependent and only a few bands
are taken into account. %
Hence, a first-principles description %
without adjustable parameters is highly desirable. 
The GW approximation (GWA) \cite{Hedin,Aryasetiawan} has been successfully
applied to calculations of excited state properties of a wide range of
materials, thus the GWA may be appropriate for this problem. %
However, due to a system size of VO$_{2}$ (monoclinic VO$_{2}$ has 12 atoms
in a unit cell), a direct application of the GW method to this system has
been limited. %
Continenza \textit{et al}. applied a simple model GW scheme to insulating VO$%
_{2}$ \cite{Continenza}, and succeeded in reproducing the band gap. In their
approach, the self-energy is approximated as a \textit{static} non-local
potential in which the experimental value of the dielectric constant is
used. Very recently Gatti \emph{et al}. performed a quasiparticle
calculation with a simplified self-consistent GW scheme\cite{Gatti}; they
first carried out a self-consistent calculation within Hedin's Coulomb hole
and screened exchange (COHSEX) approximation \cite{Hedin} and used the
resulting self-consistent one-particle wave functions and energies as an
input for a one-shot GW calculation. Using this procedure they also obtained
a band gap in good agreement with experiment. 

In the present work, we also perform GW calculations in order to gain better
understanding of electron correlation effects in VO$_{2}$, with special
emphasis on the influence of the diagonal self-energy and the role of
dynamical effect in the self-energy, going beyond a model or static
treatment of the self-energy. We find that the self-energy is in fact
strongly energy dependent and it affects significantly the one-electron
excitation spectrum. We also find that the off-diagonal self-energy can have
a large influence on the quasiparticle band structure. Our result reproduces
both the metallic and insulating VO$_{2}$. 

\section{Method}

In the GW method \cite{Hedin,Aryasetiawan}, the self-energy $\Sigma $ is
written as the product of the one-particle Green's function $G$ and the
screened Coulomb interaction $W$ as 
\begin{equation}
\Sigma (\mathbf{r},\mathbf{r}^{\prime },\omega )=\frac{i}{2\pi }\int G(%
\mathbf{r},\mathbf{r}^{\prime },\omega +\omega ^{\prime })W(\mathbf{r},%
\mathbf{r}^{\prime },\omega ^{\prime })d\omega ^{\prime }.  \label{eq:gw}
\end{equation}%
Here $W$ is calculated within the random phase approximation (RPA). 
Quasiparticle wavefunctions $\{f_{\mathbf{k}\nu }\}$ and energies $%
\{\varepsilon _{\mathbf{k}\nu }^{\mathrm{GW}}\}$ satisfy the following
equation 
\begin{eqnarray}
\bigl [ &-&\frac{1}{2}\nabla ^{2}+v_{\mathrm{ext}}(\mathbf{r})+v_{H}(\mathbf{%
r})\bigr ]f_{\mathbf{k}\nu }(\mathbf{r})  \nonumber  \label{eq:qpeq} \\
&+&\int \Sigma (\mathbf{r},\mathbf{r}^{\prime },\varepsilon _{\mathbf{k}\nu
}^{\mathrm{GW}})f_{\mathbf{k}\nu }(\mathbf{r}^{\prime })d^{3}r^{\prime
}=\varepsilon _{\mathbf{k}\nu }^{\mathrm{GW}}f_{\mathbf{k}\nu }(\mathbf{r}),
\end{eqnarray}%
where $v_{\mathrm{ext}}$ and $v_{H}$ are the external and Hartree potential,
respectively. Usually Eq.~(\ref{eq:qpeq}) is solved in a non-self-consistent
way with two further approximations: first, %
the self-energy correction from the LDA exchange-correlation potential $v_{%
\mathrm{xc}}^{\mathrm{LDA}}$ is assumed to be diagonal with respect to LDA
wavefunctions, $\Delta \Sigma _{\mu \nu }(\mathbf{k},\omega )=\langle \psi _{%
\mathbf{k}\mu }^{\mathrm{LDA}}|\hat{\Sigma}(\omega )-\hat{v}_{\mathrm{xc}}^{%
\mathrm{LDA}}|\psi _{\mathbf{k}\nu }^{\mathrm{LDA}}\rangle =\delta _{\mu \nu
}\Delta \Sigma _{\nu \nu }(\mathbf{k},\omega )$ . Second, the frequency
dependence of the self-energy is simplified by the linearization around the
LDA eigenenergies, 
\begin{eqnarray}
\Sigma _{\nu \nu }(\mathbf{k},\omega ) &\approx &\Sigma _{\nu \nu }(\mathbf{k%
},\varepsilon _{\mathbf{k}\nu }^{\mathrm{LDA}})  \nonumber
\label{eq:linearized} \\
&&+\frac{\partial \Sigma _{\nu \nu }}{\partial \omega }\bigr|_{\omega
=\varepsilon _{\mathbf{k}\nu }^{\mathrm{LDA}}}(\omega -\varepsilon _{\mathbf{%
k}\nu }^{\mathrm{LDA}}).
\end{eqnarray}%
Then the quasiparticle energies are evaluated as the first-order correction
to the Kohn-Sham eigenvalues 
\begin{equation}
\varepsilon _{\mathbf{k}\nu }^{\mathrm{GW}}=\varepsilon _{\mathbf{k}\nu }^{%
\mathrm{LDA}}+Z_{\mathbf{k}\nu }\Delta \Sigma _{\nu \nu }(\mathbf{k}%
,\varepsilon _{\mathbf{k}\nu }^{\mathrm{LDA}}),  \label{eq:linearizedegw}
\end{equation}%
where 
$Z_{\mathbf{k}\nu }=1-\frac{\partial \Sigma _{\nu \nu }(\omega )}{\partial
\omega }\bigr|_{\omega =\varepsilon _{\mathbf{k}\nu }^{\mathrm{LDA}}}$ is
the renormalization factor.

The reason why the above two approximations have been so successful for
simple semiconductors \cite{Hybertsen,Godby} is that in these systems the
self-energy shows a smooth linear behavior around the Fermi energy, and the
quasiparticle wavefunction and LDA wavefunction are almost identical. 
Since the validity of these two approximations is not clear for correlated
system like VO$_{2}$, in this work we perform calculations without using
these approximations; a frequency dependence of the self-energy is
explicitly taken into account, and the off-diagonal elements of the
self-energy (in the Kohn-Sham basis) are included properly in the
calculation. %

The quasiparticle energies are calculated by solving the following equation: 
\begin{equation}  \label{eq:fullqpeq}
\det \bigl [ (\omega - \varepsilon^{\mathrm{LDA}}_{\mathbf{k}%
\nu})\delta_{\mu\nu} -\Re \, \Delta\Sigma_{\mu\nu}(\mathbf{k},\omega) \bigr
] = 0.
\end{equation}
Here $\Re$ means the hermitian part (i.e. $\Re \, A = \frac{1}{2}%
(A+A^{\dagger})$). %
When off-diagonal elements of $\Delta\Sigma(\mathbf{k},\omega)$ are not
negligible, we calculate the roots of Eq.~(\ref{eq:fullqpeq}) by using the
following linearization scheme: first we calculate $\Delta\Sigma(\mathbf{k}%
,\omega)$ on uniform frequency meshes $\omega_{j}=\omega_{1}+\Delta\omega
(j-1)$. Then in each region $\omega_{j}\le \omega \le \omega_{j+1}$, we
linearly interpolate $\Delta\Sigma_{\mu\nu}(\mathbf{k},\omega)$ as 
\begin{eqnarray}
\Delta\Sigma_{\mu\nu}(\mathbf{k},\omega) &=& \frac{\omega_{j+1}-\omega}{%
\omega_{j+1}-\omega_{j}} \Delta\Sigma_{\mu\nu}(\mathbf{k},\omega_{j}) 
\nonumber \\
&&+ \frac{\omega-\omega_{j}}{\omega_{j+1}-\omega_{j}} \Delta\Sigma_{\mu\nu}(%
\mathbf{k},\omega_{j+1}) .
\end{eqnarray}
By substituting the above approximation for $\Delta\Sigma_{\mu\nu}(\mathbf{k}%
,\omega)$ into Eq.~(\ref{eq:fullqpeq}), the roots of Eq.~(\ref{eq:fullqpeq}) 
are easily calculated by solving the generalized eigenvalue problem. We note
that unlike the usual linearlization (Eq.~(\ref{eq:linearized})), this
scheme is exact in the limit $\Delta\omega \to 0$.

Our calculation is based on full-potential LMTO basis\cite{Methfessel}. 
The product-basis technique is used\cite{Aryasetiawan1994}, and the
frequency integral in Eq.~(\ref{eq:gw}) is numerically carried out along the
imaginary axis with the contributions from the poles of the Green function
added\cite{Aryasetiawan2000}. Details of the GW code are described in Ref.~%
\onlinecite{Kotani}. Vanadium $3s,3p$ electrons are treated as valence
electrons, and $151$ unoccupied bands per V$_{2}$O$_{4}$ are used to compute 
$G$ and $W$. To check the convergence of our calculation, we also performed
calculations with $100$ unoccupied bands per V$_{2}$O$_{4}$, and found that
the calculated band gaps in the insulating state differ only by about $0.01$
eV. For the sampling of the Brillouin zone, $6 \times 6 \times 6$ ($4 \times
4 \times 4$) Monkhorst-Pack grid \cite{Monkhorst} is used for R (M$_{1}$)
structure. In both phases, experimental lattice parameters are used\cite%
{McWhan,Longo}.

\section{Results and discussion}

\subsection{Metallic tetragonal VO$_{2}$}

\begin{figure}[tbp]
\includegraphics[width=8cm,clip]{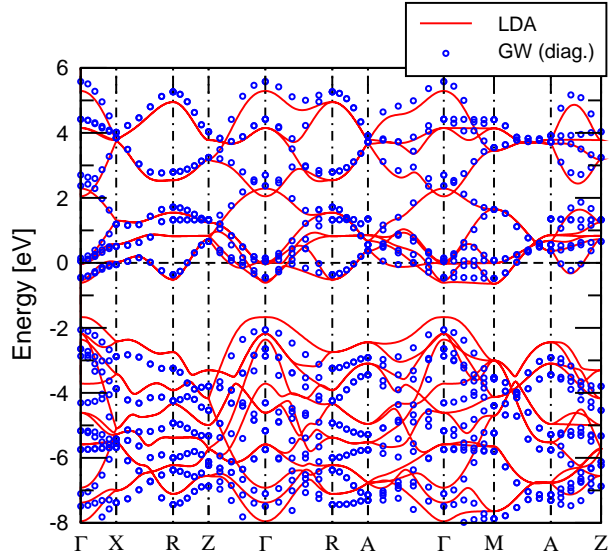}
\caption{Band structure of metalic VO$_{2}$ calculated with the LDA (solid
lines) and GW with only the diagonal self-energy (circles).}
\label{fig:vo2rband}
\end{figure}

Figure \ref{fig:vo2rband} shows the band structure of metallic VO$_{2}$
calculated within the LDA and the GWA. We find that the change in
quasiparticle energy due to the off-diagonal self-energy is negligible so
that we consider only the diagonal self-energy. The broad oxygen $2p$ band
lies from $-8$ eV to $-2$ eV, and a separation between $t_{2g}$ bands (which
extends from $-1$ eV to $2$ eV) and $e_{g}$ bands (from $2$ eV to $5$ eV) is
seen. This result is in agreement with previous calculations \cite{Eyert}. 
Compared to the LDA result, the self-energy correction makes the \textit{d-p}
separation larger by about $0.6$ eV, but the bandwidth of the \textit{d}
bands does not show a significant change. We find that due to the dynamical
correlation effects, at some k-points (not seen in Fig.~\ref{fig:vo2rband})
the number of roots of Eq.(\ref{eq:fullqpeq}) becomes greater than the
number of LDA bands considered. %
In order to study the origin of this dynamical effect, in Fig.~\ref%
{fig:vo2rsigma} we plot the frequency dependence of the diagonal self-energy 
$\Delta \Sigma _{\nu \nu }(\mathbf{k},\omega )$ for $t_{2g}$ bands at X and
R points as representative points. When the off-diagonal self-energy is
neglected, the roots of the quasiparticle equation (Eq.(\ref{eq:fullqpeq}))
are given graphically by the intersections of two lines $\omega -\varepsilon
_{\mathbf{k}\nu }^{\mathrm{LDA}}$ and $\text{Re}\,\Delta \Sigma _{\nu \nu }(%
\mathbf{k},\omega )$. For these bands a noticeable peak is found at 
around $2-3$ eV in $\text{Re}\,\Delta \Sigma _{\nu \nu }(\mathbf{k},\omega )$%
, which produces the extra solutions of Eq.(\ref{eq:fullqpeq}). Noting that
in the GW approximation the imaginary part of the self-energy is related to $%
W$ as \cite{Aryasetiawan} 
\begin{eqnarray}
\func{Im}\Sigma _{\nu \nu }^{c}(\mathbf{k},\omega  &>&E_{F})=-\sum_{\mathbf{q%
}}\sum_{\mu }^{unocc}\sum_{\alpha \beta }\left\langle \psi _{\mathbf{k}\nu
}\psi _{\mathbf{q-k}\mu }|B_{\mathbf{q}\alpha }\right\rangle   \nonumber \\
&&\times \func{Im}W_{\alpha \beta }^{c}(\mathbf{q},\omega -\varepsilon _{%
\mathbf{q-k}\mu })\left\langle B_{\mathbf{q}\beta }|\psi _{\mathbf{q-k}\mu
}\psi _{\mathbf{k}\nu }\right\rangle   \nonumber \\
&&\times \theta (\omega -\varepsilon _{\mathbf{q-k}\mu }), \\
\func{Im}\Sigma _{\nu \nu }^{c}(\mathbf{k},\omega  &\leq &E_{F})=\sum_{%
\mathbf{q}}\sum_{\mu }^{occ}\sum_{\alpha \beta }\left\langle \psi _{\mathbf{k%
}\nu }\psi _{\mathbf{q-k}\mu }|B_{\mathbf{q}\alpha }\right\rangle   \nonumber
\\
&&\times \func{Im}W_{\alpha \beta }^{c}(\mathbf{q},\varepsilon _{\mathbf{q-k}%
\mu }-\omega )\left\langle B_{\mathbf{q}\beta }|\psi _{\mathbf{q-k}\mu }\psi
_{\mathbf{k}\nu }\right\rangle   \nonumber \\
&&\times \theta (\varepsilon _{\mathbf{q-k}\mu }-\omega ),
\end{eqnarray}%
where $\{B_{\mathbf{q}\alpha }\}$ is an arbitrary set of basis functions and
the superscript $c$ signifies the correlation part, this peak structure is
due to the corresponding peak in $\text{Im}\,W(\omega )$ or equivalently $%
\text{Im}\,\epsilon ^{-1}(\omega )$, where $\epsilon (\omega )$ is the
dielectric function. We actually find a strong peak in $\text{Im}\,\epsilon
^{-1}(\omega )$ around $2$ eV, which may be interpreted as a sub-plasmon
arising from strong transitions from the narrow $a_{1g}$ band to empty
states just above the Fermi level. %
It is interesting to note that the peak in $\func{Im}\Sigma $ above the
Fermi level is much stronger than the peak below the Fermi level. This
difference in the strength of the peaks most likely originates from the
matrix elements, as can be clearly seen in the above expression for $\func{Im%
}\Sigma $.

\begin{figure}[tbp]
\includegraphics[width=7cm,clip]{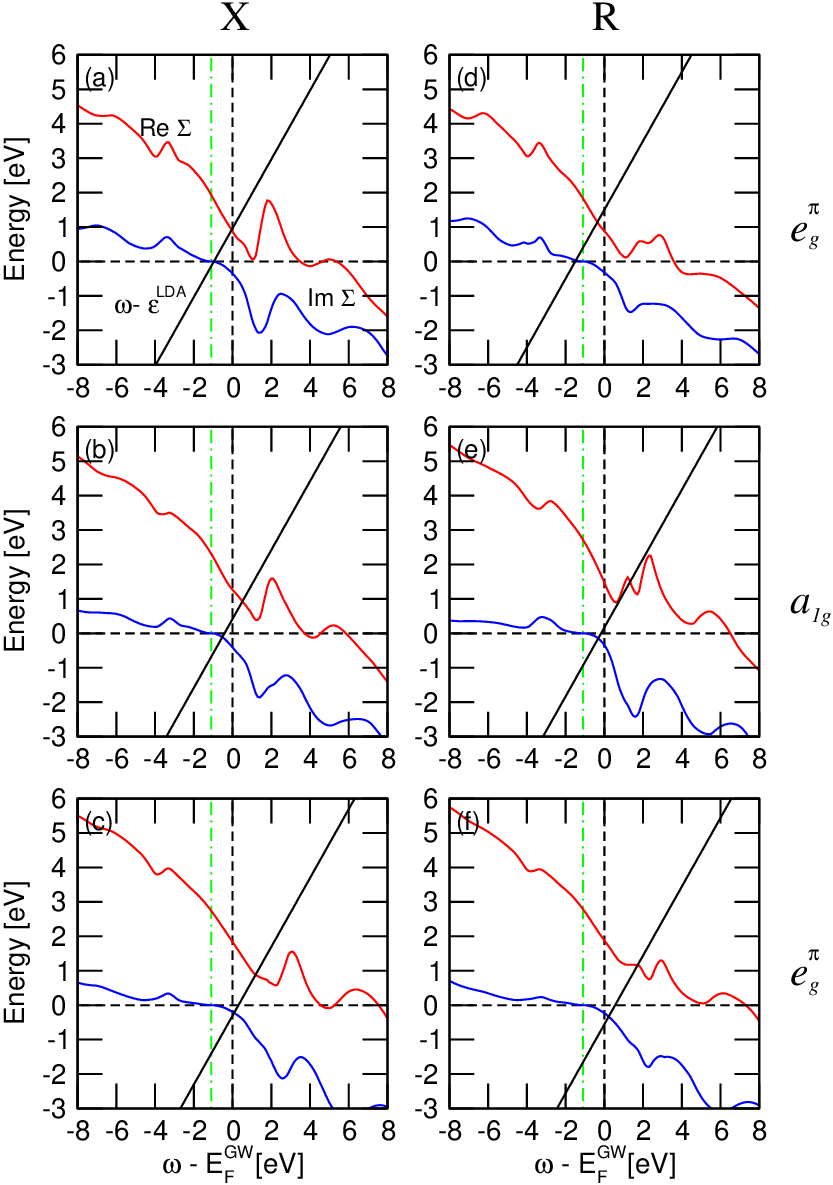}
\caption{Diagonal self-energy $\Delta \Sigma _{\protect\nu \protect\nu }(%
\mathbf{k},\protect\omega )$ for three t$_{2g}$ bands of metallic VO$_{2}$
at the $X$ point ((a)-(c)) and $R$ point ((d)-(f)). In (b) and (e) the
wavefunctions have mainly $a_{1g}$ character, and in (a)(c)(d)(f) the
wavefunctions have mainly $e_{g}^{\protect\pi}$ character. Red lines: $%
\mathrm{Re}\,\Delta \Sigma _{\protect\nu \protect\nu }(\mathbf{k},\protect%
\omega )$. Blue lines: $\mathrm{Im}\,\Delta \Sigma _{\protect\nu \protect\nu %
}(\mathbf{k},\protect\omega )$. Black solid lines: $\protect\omega -\protect%
\varepsilon _{\mathbf{k}\protect\nu }^{\mathrm{LDA}}$. The origin of the
horizontal axes is set to the renormalized Fermi energy $E_{F}^{\mathrm{GW}}$%
, and $E_{F}^{\mathrm{LDA}}$ is shown as vertical dash-dotted lines. }
\label{fig:vo2rsigma}
\end{figure}

In the metallic phase the orbital dependence of the self-energy for these $%
t_{2g}$ orbitals is small, which reflects the weak orbital polarization in
this phase. We also find that the self-energy shows some k-dependence; for
the X point (Fig.~\ref{fig:vo2rsigma}(a)-(c)) and other k points that lies
on the plane $k_{z}=0$, where $k_{z}$ is a reciprocal vector parallel to the 
$c$ axis, the self-energy shows similar, or ``isotropic'' behavior, while
for the R point (Fig.~\ref{fig:vo2rsigma}(d)-(f)) and other k points that
lie on $k_{z}=\frac{\pi}{c}$ plane, the self-energy is somewhat anisotropic
and a peak around $2$ eV is more pronounced for $a_{1g}$ (Fig.~\ref%
{fig:vo2rsigma}(e)) than $e_{g}^{\pi}$ (Fig.~\ref{fig:vo2rsigma}(d),(f)).

\begin{figure}[tbp]
\includegraphics[width=6cm,clip]{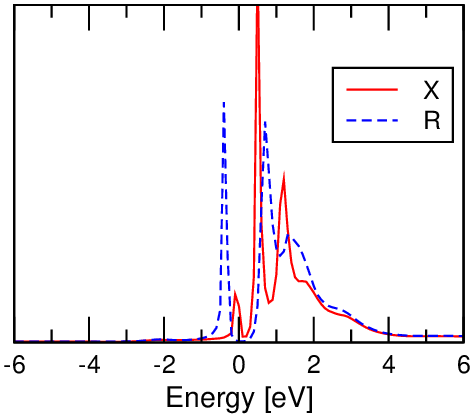}
\caption{Spectral function $A(\mathbf{k},\protect\omega )$ for t$_{2g}$
bands of metallic VO$_{2}$ at the $X$ point (solid line) and $R$ point
(dashed line).}
\label{fig:vo2rakw}
\end{figure}

To see the consequence of the sub-plasmon peak, we plot the diagonal
spectral function $A(\mathbf{k},\omega )=\sum_{\nu }\frac{1}{\pi }\bigl |%
\mathrm{Im}\frac{1}{\omega -\varepsilon _{\mathbf{k}\nu }^{\mathrm{LDA}%
}-\Delta \Sigma _{\nu \nu }(\mathbf{k},\omega )}\bigr |$ for $t_{2g}$ bands
at $X$ and $R$ %
in Fig.~\ref{fig:vo2rakw}. In calculating $A(\mathbf{k},\omega )$, since the
self-energy is calculated with the unshifted LDA Fermi energy, we shift the
frequency dependence of $\mathrm{Im}\Sigma $ as $\mathrm{Im}\Sigma (\omega
)\rightarrow \mathrm{Im}\Sigma (\omega -\Delta E_{F})$, where $\Delta
E_{F}=E_{F}^{\mathrm{GW}}-E_{F}^{\mathrm{LDA}}$, so as to reproduce the
small inverse lifetime of quasiparticles around the Fermi energy. Close to
the Fermi energy there are sharp quasiparticle peaks, and the
sub-plasmon-originated peak at around $2-3$ eV in the self-energy yields a
weak satellite structure. The renormalization factor $Z_{\mathbf{k}\nu }$ of 
$t_{2g}$ bands is about $0.5$, which indicates strong transfer of the
spectra weight to the incoherent part. In the DMFT calculation by Liebsch 
\textit{et al.} \cite{Liebsch} and the cluster DMFT calculation by Biermann 
\textit{et al.} \cite{Biermann} the satellite structure is found also below
the quasiparticle peak at around $-1.5$ eV, which is also observed in the
recent photoemission spectra \cite{Koethe}. That satellite, regarded as the
lower Hubbard band, is not seen in our calculation.

The self-energy of real systems differs from that of the Hubbard model in
the following fashions. The energy scale of the self-energy of the Hubbard
model is determined by $U$, which is of the order of a few eV, whereas the
energy scale of the self-energy of the real systems is determined by the
bare Coulomb interaction, which is typically one order of magnitude larger
than the Hubbard $U$. As a result, $\func{Im}\Sigma $ of the Hubbard model
decays to zero after a few eV, above and below the Fermi level, whereas $%
\func{Im}\Sigma $ of the real systems decays to zero at energies larger than
the plasmon energy, which is of the order of tens of eV.

As can be seen in Fig.~\ref{fig:vo2rsigma} the presence of high-energy
plasmon excitation causes $\func{Im}\Sigma $ to increase to a large value
outside the energy range of approximately -4 eV and 4 eV, in contrast to the
Hubbard model in which $\func{Im}\Sigma $ decays to zero outside this energy
range. The decay of $\func{Im}\Sigma $ enhances the satellite structure and
may actually overemphasize the strength of the satellite, as discussed in
Ref.~\onlinecite{Aryasetiawan2}. Nevertheless, experimentally there appears
to be some evidence that there exists indeed a satellite structure a few eV
below the Fermi level \cite{Koethe}. In order to reproduce that peak, one
must include the short range correlation effect beyond RPA; for this purpose
it would be very interesting to apply methods such as GW+DMFT \cite%
{Biermann2} to this system.

\subsection{Insulating monoclinic VO$_{2}$}

\begin{figure}[tbp]
\includegraphics[width=8cm,clip]{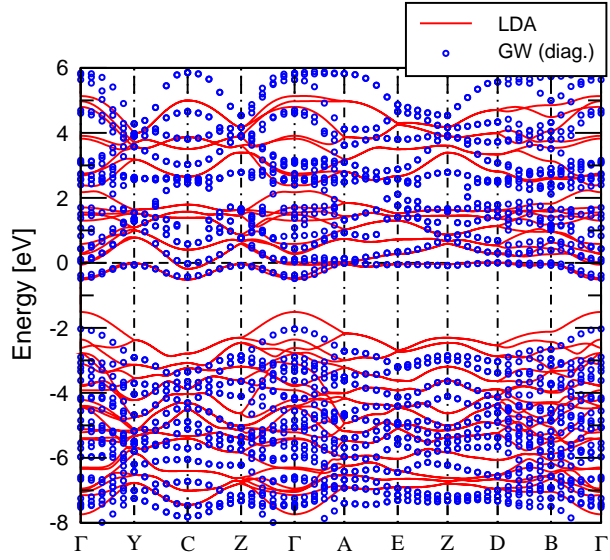}
\caption{Band structure of insulating VO$_{2}$ calculated with the LDA
(solid lines) and GW including only a diagonal part of the self-energy
(circles). The symmetry labels are the same as those in Ref.~ 
\onlinecite{Eyert}.}
\label{fig:vo2m1band}
\end{figure}

\begin{figure}[tbp]
\includegraphics[width=8cm,clip]{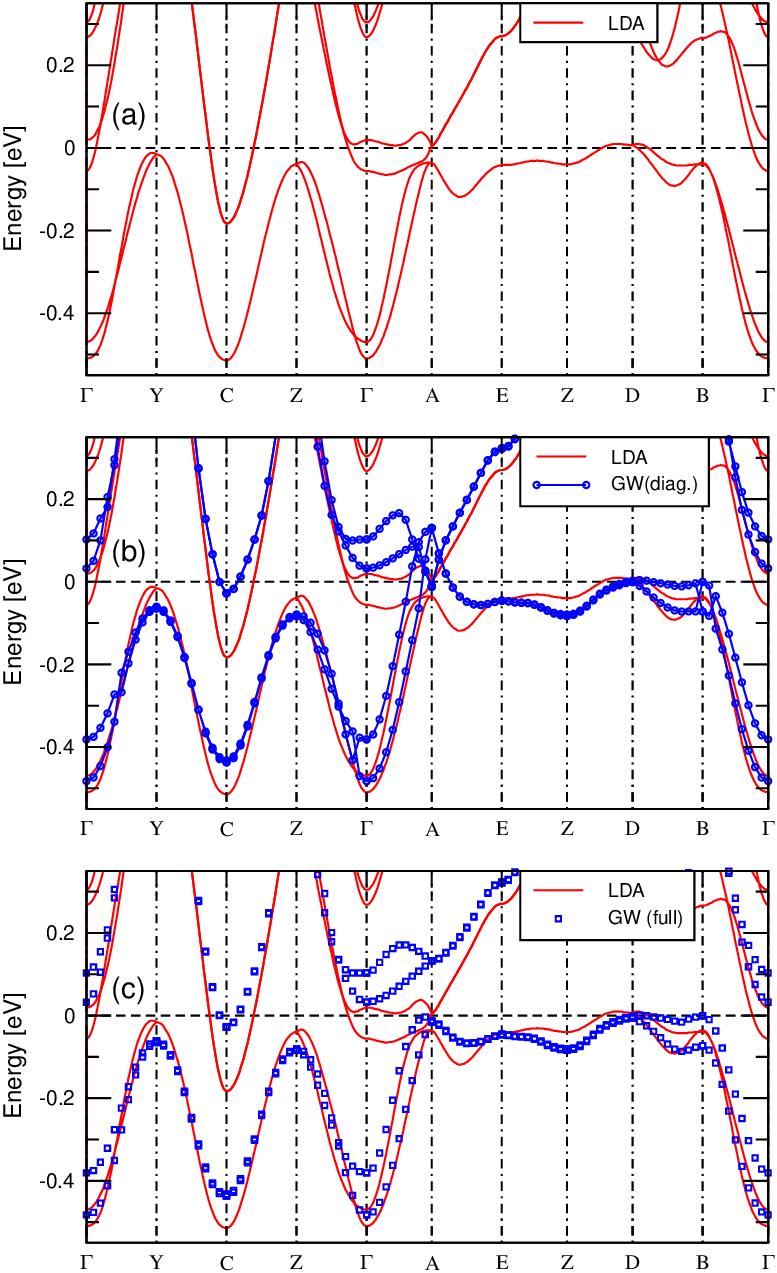}
\caption{Band structure of insulating VO$_{2}$ near the Fermi energy
calculated with (a)LDA, (b)GW including only a diagonal part of the
self-energy, (c)GW including the off-diagonal self-energy.}
\label{fig:vo2m1band2}
\end{figure}

Figure \ref{fig:vo2m1band} shows the band structure of insulating monoclinic
VO$_{2}$ calculated within the LDA and the GWA with only the diagonal part
of the self-energy. As in the metallic phase we see the separation of the $%
t_{2g}$ and $e_{g}$ bands due to the octahedral crystal field and the
downward shift of O $2p$ bands, but in this phase the occupied part of V $3d$
states is mainly of $a_{1g}$ character, which results from the pairing of
the vanadium atoms. To evaluate the effect of the off-diagonal self-energy
to the band structure, we compare the results with and without the
off-diagonal self-energy in Fig.~\ref{fig:vo2m1band2}. In this calculation
the matrix elements of the self-energy are calculated within the V $d$
subspace. We find that inclusion of the O $2p$ bands changes the
quasiparticle energies by less than $0.1$ eV. In most bands the off-diagonal
self-energy is very small and the two results are almost the same, but a
noticeable difference is found near the Fermi energy. In the diagonal-only
GW result (Fig.~\ref{fig:vo2m1band2}(b)), an unusual band overlap around the
A point is found, which is removed when the off-diagonal self-energy is
included. This anomaly is attributed to the inefficiency of both the LDA and
1-shot GW. Around the A point the valence ($a_{1g}$) band and conduction ($%
e_{g}^{\pi }$) band are barely separated in the LDA level (Fig.~\ref%
{fig:vo2m1band2}(a)), so these bands are \textquotedblleft
fictitiously\textquotedblright\ hybridized. Since the diagonal-only GW
calculation does not change the character of wavefunctions, this
hybridization cannot be removed unless the off-diagonal self-energy is
included. Thus, the LDA wavefunctions and the quasiparticle ones obtained by
including the off-diagonal self-energy are far from identical.

At first sight the presence of bands crossing the Fermi level at around the
A point intuitively suggests that it is unlikely to open up a gap around
that point. In the case of single band crossing the Fermi level there
appears to be no choice other than an opening of a Mott gap. However, in the
case of VO$_{2}$ there are multi-bands, which still allow for an opening of
a gap within a one-particle picture by means of rearrangement of the band
occupation. This is indeed the case and it is noteworthy that the
rearrangement of the band occupation around the A point is already obtained
in the one-shot GW calculation, provided the off-diagonal self-energy is
taken into account.

We thus observe two important ingredients in gap opening in narrow band
materials where entangled bands cross the Fermi level: First, the
off-diagonal self-energy is crucial in "dehybridizing" the bands and
rearranging the band occupation, and second, the modification of the
one-particle energies decreases the screening and hence enhances the gap
opening \cite{miyake00,sakuma08} in the course of self-consistency. This is
in contrast to the conventional semiconductors where the gap opening, in the
case of overlapping valence and conduction bands, is simply affected by
shifting the bands, i.e., by the second mechanism.

\begin{figure}[tbp]
\includegraphics[width=7cm,clip]{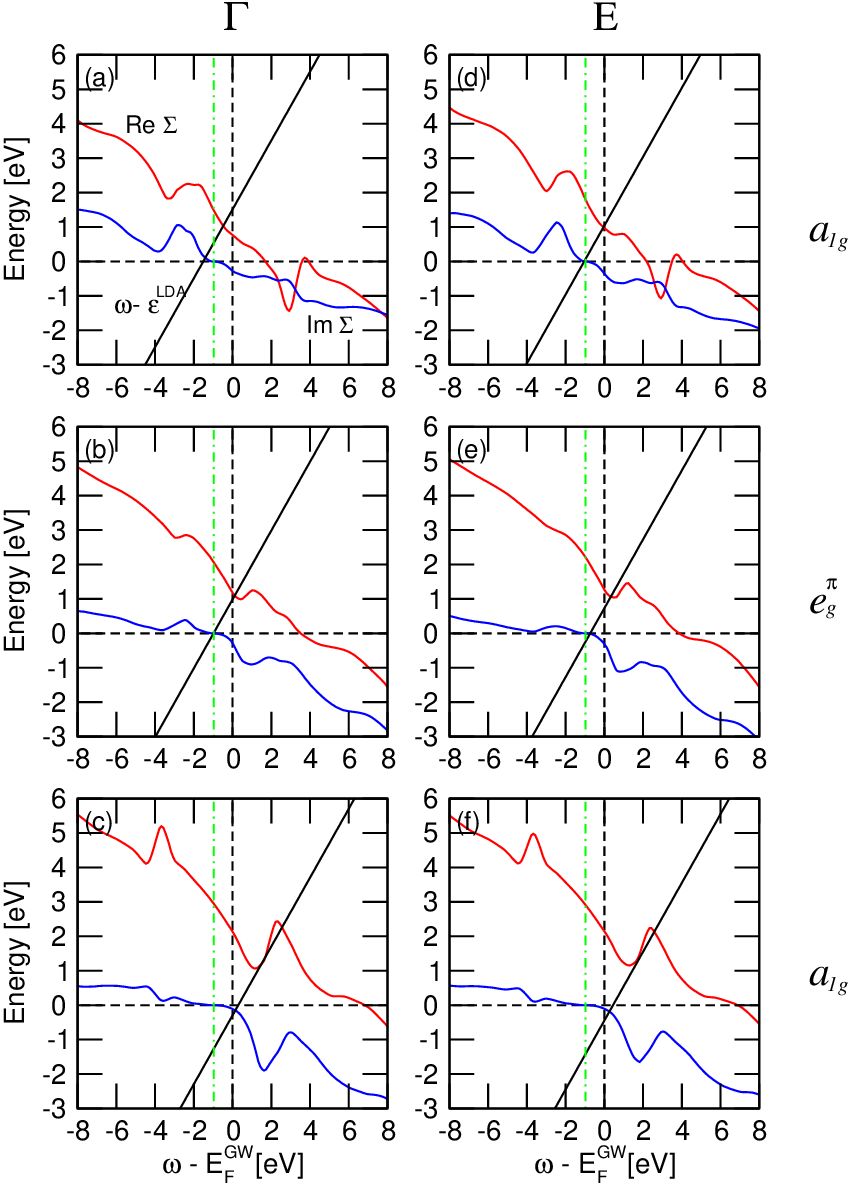}
\caption{Diagonal self-energy $\Delta\Sigma_{\protect\nu\protect\nu}(\mathbf{%
k},\protect\omega)$ of insulating VO$_{2}$ for t$_{2g}$ bands in the 1-shot
GW calculation at the $\Gamma$ point (a-c) and $E$ point (d-f). (a) and (d)
: bonding $a_{1g}$ bands. (b) and (e) : $e_{g}^{\protect\pi}$ bands. (c) and
(f) : anti-bonding $a_{1g}$ bands. Red lines: $\mathrm{Re}\,\Delta\Sigma_{%
\protect\nu\protect\nu}(\mathbf{k},\protect\omega)$. Blue lines: $\mathrm{Im}%
\,\Delta\Sigma_{\protect\nu\protect\nu}(\mathbf{k},\protect\omega)$. Black
solid lines: $\protect\omega-\protect\varepsilon_{\mathbf{k}\protect\nu}^{%
\mathrm{LDA}}$. The origin of the horizontal axes is set to the renormalized
Fermi energy $E_{F}^{\mathrm{GW}}$, and $E_{F}^{\mathrm{LDA}}$ is shown as
vertical dash-dotted lines.}
\label{fig:vo2m1sigma}
\end{figure}

We also find that due to the non-linear behavior of the self-energy, it is
crucial to take into account the frequency dependence of the self-energy
explicitly in calculating $\{\varepsilon^{\mathrm{GW}}_{\mathbf{k}\nu}\}$.
The results with and without the linearization are compared in Fig.~\ref%
{fig:vo2m1bandcompare}. The usual linearized approach (Eq.~(\ref%
{eq:linearizedegw})), which is used in most 1-shot GW calculations, causes
an error of as much as $0.2$ eV for the $a_{1g}$ band, which leads to the
underestimation of the valence band width and hence overestimation of the
band gap. Figure ~\ref{fig:vo2m1sigma} (a-c) shows the diagonal self-energy
for the insulating phase at $\Gamma $ and E for three bands: bonding and
antibonding $a_{1g}$ and $e_{g}^{\pi }$. The self-energy for other k-points
shows similar behavior. A nonlinear behavior of the self-energy is clearly
seen from the figure, which explains the failure of usual linearization
scheme. Owing to the lattice distortion, the self-energy for these $t_{2g}$
bands shows strong orbital dependence, in contrast to the metallic phase;
peak structures in $\func{Re}\Sigma $ around $-4$ eV and $1-3$ eV arising
from strong transition from the narrow $a_{1g}$ band to empty states just
above the Fermi level are more prominent for $a_{1g}$ bands. This peak
yields extra solutions of Eq.~(\ref{eq:fullqpeq}), which are seen as flat
bands around $2.5$ eV %
in Fig.~\ref{fig:vo2m1band}. The renormalization factor for these bands is
about $0.5-0.6$.

In our 1-shot GW result, even if the off-diagonal self-energy is included
there still exists a small indirect band overlap of $0.02$ eV in contrast to
the experimental band gap of $0.6$ eV. Furthermore, the bonding-antibonding
splitting of $a_{1g}$ bands is less than $2$ eV, noticeably smaller than
experiment\cite{Koethe}. The present calculation uses the LDA wavefunctions
and eigenenergies to construct $G$ and $W$. Thus, the failure to reproduce
the experimental gap may be due to the poorness of the starting states, 
and a calculation starting from a better mean-field solution will be needed
to test the initial state dependence. Indeed, very recently Gatti \emph{et al%
} attempted such a calculation \cite{Gatti}. 
They first performed a static COHSEX calculation self-consistently, and
obtained an insulating phase. The result is used as the starting Hamiltonian
of a subsequent one-shot GW calculation, which yields a final band gap in
good agreement with experiment. %
They found that without modification of the LDA wave functions, the gap is
not opened. 

Since a self-consistent GW scheme \cite{Holm,Ku,Kotani,Bruneval,Shishkin} is
not yet well-established\cite{Delaney,Ku_reply}, it would be meaningful to
study the effect of initial states \cite{miyake06} and self-consistency. The
problem is how to construct a one-particle Hamiltonian whose eigenfunctions
and energies well represent the quasiparticle wavefunctions and energies.
One possible way is a so-called quasiparticle self-consistent scheme (QSGW) 
proposed in Ref.~\onlinecite{Faleev}, where the exchange and correlation
part of the one-particle Hamiltonian is replaced by a static non-local
exchange-correlation potential constructed using the following formula:

\begin{equation}
\langle \psi^{\mathrm{LDA}}_{\mathbf{k}\mu} | \hat{v}_{\mathrm{xc}} | \psi^{%
\mathrm{LDA}}_{\mathbf{k}\nu} \rangle =\frac{\Re}{2} [\Sigma _{\mu\nu}(%
\mathbf{k},\varepsilon^{\mathrm{GW}}_{\mathbf{k}\mu}) +\Sigma _{\nu\mu}(%
\mathbf{k},\varepsilon^{\mathrm{GW}}_{\mathbf{k}\nu})]  \label{QSGW}
\end{equation}%
with the self-energy $\Sigma $ calculated in the GWA. The reason for
introducing such a recipe is that the self-energy is energy dependent. For
the diagonal elements ($\mu=\nu$) no ambiguities arise in choosing the
energy but for the off-diagonal elements the choice of the energy is
ambiguous. The formula, however, still awaits theoretical justification.

As another choice, we propose the following procedure. The quasiparticle
wavefunctions $\{\Psi _{\mathbf{k}\nu }\}$, obtained as solutions of Eq. (%
\ref{eq:qpeq}) with $\func{Im}\Sigma $ neglected, are in general not
orthogonal and therefore unsuitable as an input for a GW calculation. 
We envisage that a natural and physically motivated quasiparticle
Hamiltonian is%
\begin{equation}
H_{\mathrm{QP}}=\sum_{\mathbf{k}\nu }\left\vert \Psi _{\mathbf{k}\nu
}\right\rangle \varepsilon _{\mathbf{k}\nu }^{\mathrm{GW}}\left\langle \Psi
_{\mathbf{k}\nu }\right\vert ,  \label{HQP}
\end{equation}%
which may be viewed as a series of quasiparticle modes (QPM). This
quasiparticle Hamiltonian is evidently Hermitian and %
ambiguities in choosing the energy in the self-energy do not arise. Upon
diagonalizing $H_{\mathrm{QP}}$ we obtain an orthonormal set of
wavefunctions with the corresponding eigenvalues. In most cases the
quasiparticle wavefunctions are almost orthogonal, thus the solutions of
Eq.~(\ref{HQP}) should be very close to the original quasiparticle states.

\begin{figure}[tbp]
\includegraphics[width=8cm,clip]{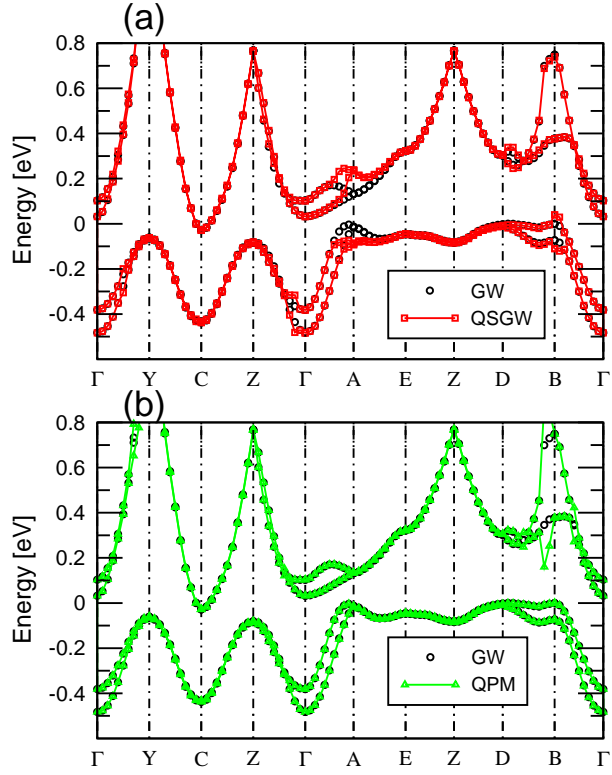}
\caption{Band structure of insulating VO$_{2}$ calculated with (a)
GW(circles) and QSGW(squares) and (b) GW(circles) and QPM(triangles). }
\label{fig:qpaandb}
\end{figure}

To convince ourselves that our QPM approximation is physically sound, we
compare in Fig.~\ref{fig:qpaandb} the quasiparticle band structures
generated from the solutions of Eq.~(\ref{eq:fullqpeq}) and the QPM
approximation (Eq.~(\ref{HQP})) as well as the QSGW (Eq.~(\ref{QSGW})). In
this calculation, the self-energy $\Sigma $ is calculated with LDA
eigenstates (i.e., non-self-consistent calculation). As can be seen the QPM
band structure agree with the true quasiparticle band structure well within
the accuracy of the calculations whereas the band structure calculated using
Eq.~(\ref{QSGW}) shows a noticeable deviation from the true quasiparticle
band structure around the A point. This deviation originates from the
approximated form of the off-diagonal part of the effective
exchange-correlation potential. Thus, the quasiparticle energy can be
sensitive to the choice of the energy in the self-energy. On the other hand,
the QPM approximation is free from ambiguity in choosing the energy argument
in the self-energy and the off-diagonal self-energy is properly taken into
account implicitly via the quasiparticle wavefunction (Eq.~(\ref{HQP})).

A self-consistent scheme is now at our disposal. We start by performing a GW
calculation using the LDA band structure as an input. The resulting
self-energy is then used to construct the quasiparticle wavefunctions and
energies by solving Eq.~(\ref{eq:fullqpeq}). A new set of wavefunctions and
energies are obtained within the QPM approximation in Eq.~(\ref{HQP}) and
used to perform the next GW calculation. The iteration continues until the
quasiparticle wavefunctions and energies converge. For the case of VO$_{2}$
that we are considering, this is a very cumbersome task due to the large
system size. We have observed, however, the following result: Starting from
the metallic LDA band structure, the quasiparticle band structure obtained
by solving Eq.~(\ref{eq:fullqpeq}) acquires a (direct) gap already in the
first iteration. The self-energy is assumed to be diagonal except within the
V $d$ subspace. It is important to note that the gap around the A point is
not opened up if the off-diagonal elements of the self-energy within the t$%
_{2g}$ subspace are not taken into account. This suggests that the
rearrangement of the wavefunctions already takes place in the first
iteration and after the first iteration we may then keep the same
wavefunctions. Thus in the subsequent iterations we simulate the
self-consistency by uniformly shifting the conduction band quasiparticle
energies as $E_{\mathbf{k}c}\rightarrow E_{\mathbf{k}c}+\Delta $. We tested
this simplified self-consistent scheme for simple semiconductors and found
that it yields a result very close to that of the full QPM self-consistent
calculation. In the calculation of VO$_{2}$, we start the calculation from
the gapped state by adding an initial shift to the LDA conduction levels.
This process is necessary to better simulate the insulating state in
constructing the self-energy matrix. We performed the calculations with
several values for the initial shift, and found that the final result is not
sensitive to the value of the initial shift.

The result of the gaps as a function of the shift is shown in Fig.~\ref%
{fig:gapscissors}. Two indirect gaps are shown, one between points B and C
and the other between D and C. The GW gaps increase almost linearly with the
shift, and a self-consistent value of $0.6$ eV is found within the QPM
approximation. This result differs considerably from the result obtained by
assuming that the self-energy is diagonal, i.e., the wavefunctions are given
by those of the LDA. %
Our results %
show that the GWA within the quasiparticle concept is sensitive to the
treatment of the off-diagonal elements of the self-energy, which modify the
LDA wavefunctions. Although a proper definition of effective quasiparticle
Hamiltonian is debatable \cite{Gatti,Shishkin,Faleev}, the obtained $0.6$ eV
gap in our calculation, which happens to be in good agreement with the
experimental value, is an indication that our QPM scheme could furnish a
suitable way of constructing a quasiparticle Hamiltonian. 

Our result makes us reconsider the electronic state of the paramagnetic
insulating VO$_{2}$; in Peierls description, the bonding-antibonding
separation of $a_{1g}$ bands is determined only by the hybridization between
two $a_{1g}$ bands, and both spin up and down states are assumed to occupy
the same (bonding $a_{1g}$) orbitals. However, the Coulomb interaction
between localized $d$ electrons invalidates this interpretation, as is clear
from an example of 2-site Hubbard model. Thus, the paramagnetic insulating VO%
$_{2}$ may be the result of a singlet state formed by two localized
electrons on paired vanadium atoms, which cannot be described in the
mean-field theory. This is indeed the conclusion drawn from the work of
Biermann \emph{et al }\cite{Biermann}, who performed cluster LDA+DMFT
calculations taking the vanadium dimer as a unit. However, our result along
with previous calculations \cite{Continenza, Gatti} suggest that a
one-particle description can still go a long way in describing the
electronic structure of VO$_{2}$, at least in describing the insulating gap.
Nevertheless, we think that the Peierls picture alone may not be sufficient
for a complete description of the electronic structure of the insulating VO$%
_{2}$. The missing satellites below the Fermi level, observed in
photoemission experiment, is one such indication.

\section{Conclusion}

In conclusion, we have investigated the self-energy and band structure of
the metallic and insulating phases of VO$_{2}$ within the GWA. In agreement
with the work of Gatti \emph{et al}, we have found that the GWA is able to
describe the metal-insulator transition. %
Our calculations indicate that the band gap depends sensitively on the
approximation used for the off-diagonal elements of the self-energy. Thus, a
proper way of defining the off-diagonal elements of the self-energy is
crucial. %
Using the QPM approximation and the scissor operator we have performed a
self-consistent calculation to determine the band gap, which is found to be
0.6 eV, in agreement with calculated value by Gatti \emph{et al} albeit
using different methods. 
Our results within the QPM approximation suggests that the GWA can treat
correlation effects in VO$_{2}$ and the opening of the gap supports the
Peierls picture of Wentzcovitch \emph{et al}. It does not seem to be
necessary to perform calculations with broken spin symmetry
(antiferromagnetic structure) as commonly done in order to open up a gap 
\cite{Korotin}.

In both the metallic and insulating phases a satellite feature exists above
the Fermi energy but not below, in contrast to the spectra calculated within
the LDA+DMFT scheme. Since the experimental photoemission spectrum displays
satellite features above and below the Fermi level, usually interpreted as
the upper and lower Hubbard bands, we attribute the discrepancy of the GW
spectra to vertex corrections, i.e., correlations beyond the RPA, which are
apparently captured in the LDA+DMFT.

We have also found that the common procedure of calculating the
quasiparticle energy by linearization of the self-energy failed badly in the
case of VO$_{2}$. It would seem that for narrow band materials it is
important to take into account the full energy dependence of the self-energy
when calculating the quasiparticle energy.

\begin{acknowledgments}
We thank S. Biermann, J. Tomczak, L. Reining, and M. Gatti for fruitful
discussions. We also acknowledge the use of the full-potential LMTO-GW\ code
provided to us by T. Kotani and M. van Schilfgaarde. This work was supported
by Grant-in-Aid for Scientific Research from MEXT, Japan (Grant Nos.
19019013 and 19051016).
\end{acknowledgments}

\newpage 

\begin{figure}[tbp]
\includegraphics[width=8cm,clip]{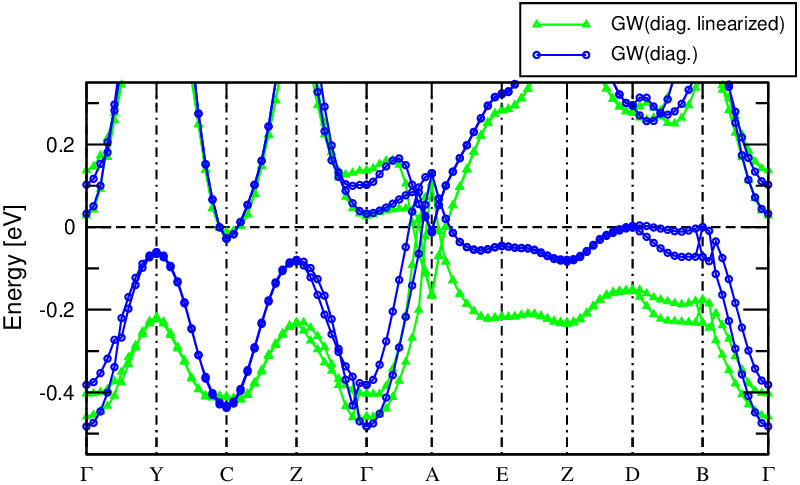}
\caption{GW band structure of insulating VO$_{2}$. Triangles(circles)
correspond to the result with(without) the linearization of the self-energy.
Only the diagonal self-energy is included in the calculation.}
\label{fig:vo2m1bandcompare}
\end{figure}

\begin{figure}[tbp]
\includegraphics[width=8cm,clip]{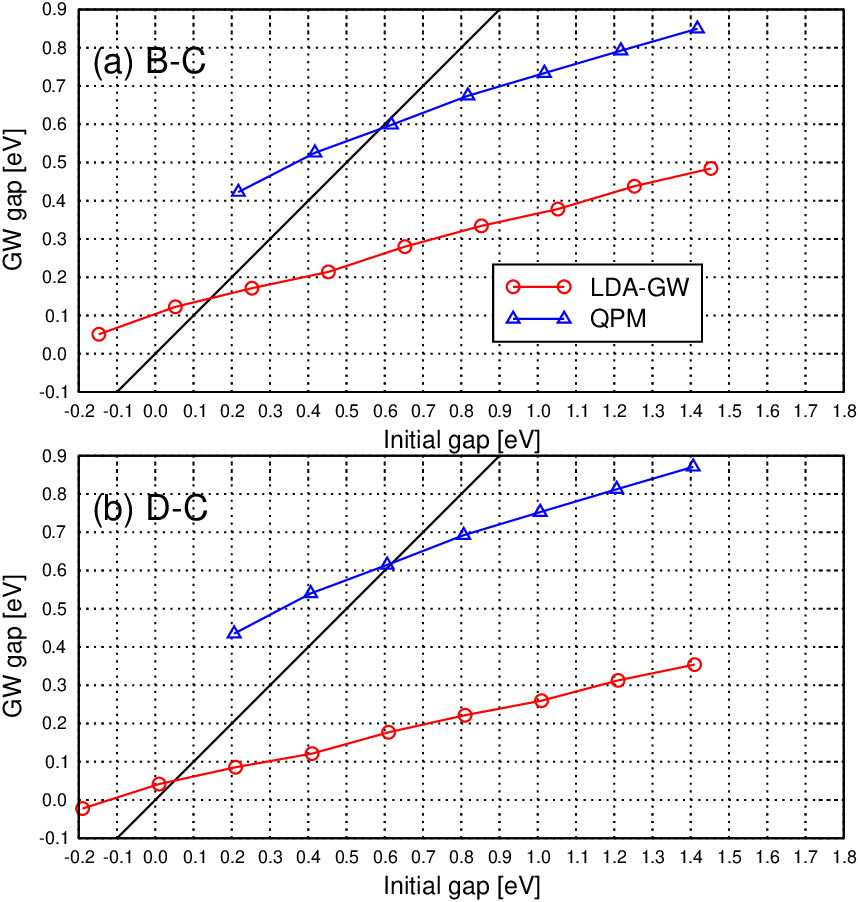}
\caption{GW indirect gap between (a)B and C and (b) D and C as a function of
the initial band gap with the self-energy calculated by using the
wavefunctions obtained within the LDA(circles) and QPM(triangles). }
\label{fig:gapscissors}
\end{figure}

\end{document}